%% file: lubashevsky_cogsci.tex
\documentclass[twocolumn,superscriptaddress]{revtex4-1}

\usepackage{natbib}

\usepackage[pdftex]{graphicx}
\usepackage{amsmath,amssymb}

\begin{document}

\title{Fractional Dynamics and Multi-Slide Model of Human Memory}
 
\author{Ihor Lubashevsky}
\email[]{i-lubash@u-aizu.ac.jp}
\affiliation{University of Aizu, Tsuruga, Ikki-machi, Aizu-wakamatsu, 965-8580 Fukushima, Japan}

\author{Bohdan Datsko}
\email[]{b_datsko@yahoo.com)}
\affiliation{Institute for Applied Problems of Mechanics and Mathematics, National Academy of Sciences,\\ Naukova st., 3b, Lviv, 79601, Ukraine}
\affiliation{University of Aizu, Tsuruga, Ikki-machi, Aizu-wakamatsu, 965-8580 Fukushima, Japan}

\begin{abstract}
We propose a single chunk model of long-term memory that combines the basic features of the ACT-R theory and the multiple trace memory architecture. The pivot point of the developed theory is a mathematical description of the creation of new memory traces caused by learning a certain fragment of information pattern and affected by the fragments of this pattern already retained by the current moment of time. Using the available psychological and physiological data these constructions are justified. The final equation governing the learning and forgetting processes is constructed in the form of the differential equation with the Caputo type fractional time derivative. Several characteristic situations of the learning (continuous and discontinuous) and forgetting processes are studied numerically. In particular, it is demonstrated that, first, the ``learning'' and ``forgetting'' exponents of the corresponding power laws of the memory fractional dynamics should be regarded as independent system parameters. Second, as far as the spacing effects are concerned, the longer the discontinuous learning process, the longer the time interval within which a subject remembers the information without its considerable lost. Besides, the latter relationship is a linear proportionality.

\end{abstract}

\maketitle

\section{Psychological and Physiological Background}

There are a number of approaches to understanding and describing processes in human mind. They belong to different levels of abstraction, ranging from neural and biochemical processes in the brain up to philosophical constructions, and study its different aspects.  In the present work we focus our attention on the phenomenological (psychological) description of human memory dealing with it as a whole, i.e., without reducing the corresponding mental functions to the real physiological processes implementing them. A review of advances made in this scope during the last decades can be found in a monograph by \citet{anderson2007can} who inspired the development of the ACT-R concept in cognitive science, a modern theory about how human cognition works.  

The ACT-R theory operates with three types of human memory, sensory, short-term, and long-term ones \cite{atkinson1968human} and accepts, in particular, the following basic postulates. 

\textit{First}, the declarative (long-term) memory is organized in chunks \cite{miller1956magical}, certain cognitive units related to some information objects. At the first approximation the learning, memorizing, and retrieval of a given object proceeds via the creation and evolution of the corresponding chunk. Naturally, chunks can interact with one another, in particular, forming larger composed chunks and, finally, their hierarchical network. The notion of chunk is general, therefore, it is rather problematic to define it more precisely, for discussion and history see, e.g., a review by \citet{anderson1998atomic}.   

\textit{Second}, each chunk individually is characterized by its strength $F$ which determines also the information retention, namely, the probability of successful retrieval of the corresponding information from a given chunk \cite{anderson1983spreading,anderson1991reflections}.
Since the classical experiments of \citet{ebbinghaus1913memory} a rather big data-set about the retention ability of human  memory has been accumulated for time scales from several minutes up to a few weeks. It has been figured out that the memory strength $F$ decays with time $t$ according to the power-law \cite{wickelgren1972trace,wickelgren1974single,anderson1991reflections,wixted1991form,wixted1997genuine}, i.e., exhibits the asymptotic behavior
\begin{equation}
F(t)\propto \frac1{t^d}\,,
\end{equation}
where the exponent $d$ is a certain constant. It should be noted that, in general, this dependence meets the second Jost's law, the increment of the strength decay becomes weaker as time goes on (see, e.g., a review by \citet{wixted2004common}). Appealing at least to the data-set collected by \citet{ebbinghaus1913memory} and analyzed by \citet{anderson1991reflections} as well as one collected and studied by \citet{wixted1991form,wixted1997genuine} the exponent $d$ seems to be rather universal and can be estimated as $d \sim 0.1-0.2$. 

The \textit{third} postulate concerns the multiple-trace arrangement of human memory. It assumes that each attempt of learning and memorizing some information fragment produces a separate trace $m$ in human memory. So the corresponding cognitive unit, a chunk, is actually a collection $\{m_i\}$ of many memory traces and its strength $F$ is the sum of their individual activation levels $\{f_i\}$ \cite{hintzman1986mtmm}
\begin{equation}\label{psyc:1}
F(t)=\sum_{m_i\in \text{Chunk}} f_i(t)\,.
\end{equation}
Evidence collected currently in physiology (see, e.g., work by \citet{yassa2013competitive} and references therein) partly supports this concept. Its implementation in physiological terms is reduced to the Multiple Trace Theory (MTT) developed by \citet{nadel1997memory} appealing to the role of the hippocampus in the encoding of new memory traces as well as the retrieval of all the previous traces, including remote ones. The preceding alternative of MTT is the Standard Model of Systems Consolidation (SMSC: \citet{squire1995retrograde}). It assumes the hippocampus to ``teach'' the cortex a memory trace strengthening the connectivity between its individual elements over time and, finally, consolidating the memory.

Recently  \citet{yassa2013competitive} have proposed a Competitive Trace Theory (CTT) combining elements SMSC and MTT. It suggests that when a memory is reactivated by a new cue, the hippocampus acts to re-instantiate the original memory traces, recombine their elements in the episodic memory, and add or subtract individual contextual features. As a result, a new memory trace overlapping with the original ones is created and ready to be stored in the neocortex. However, in contrast to MTT, CTT supposes that the memory traces are not stored in parallel but compete for representation in the neocortex. Two relative phenomena occur here: consolidation and decontextualization. First, overlapping features in the memories should not compete for representation and thus are strengthened, i.e., consolidated. Second, non-overlapping features should compete with one another resulting in mutual inhibition and, as a result, memories become decontextualized. \citet{nadel1998hippocampal} proposed the reactivation of memory traces to strengthen also the links between the traces too. The concept of such a multi-trace consolidation can be regarded as the \textit{fourth} postulate of the ACT-R theory.

As the \textit{fifth} postulate we note the following. The hippocampus is involved in the ``reconstruction'' rather than the ``retrieval'' of the memory. So, as CTT states, new memory traces are only partially but not completely overlap with the original traces. It is due to the hippocampus capability of supporting rapid encoding of unique experiences by orthogonalizing incoming inputs such that their interference is minimized, which is termed pattern separation; the available evidence for this feature was recently discussed by \citet{yassa2011pattern}. This pattern separation together with the corresponding pattern completion via creating new memory traces endows our episodic memory system with richness, associativity, and flexibility \cite{yassa2011pattern}.

\textit{Finally}, the ACT-R theory accepts an important generalization about expansion~\eqref{psyc:1}. It assumes that the individual activation levels $f_i(t)$ of memory traces decreases with time $t$ also according to the power law and, after formation, their individual dynamics is mutually independent. Thereby the strength $F(t)$ of the corresponding chunk is the superposition \cite{anderson1983spreading,anderson1991reflections}
\begin{equation}\label{psyc:2}
F(t)=\sum_{m_i\in \text{Chunk}} \frac{C_i}{(t-t_i)^d}\,,
\end{equation}
where $\{C_i\}$ are some constants and $t_i$ is the time moment when the chunk $m_i$ was created. It should be noted that expression~\eqref{psyc:2} does not take into account the chunk interaction. 

Our following constructions will be based on these postulates. In the present work we will confine our consideration to the dynamics of a single chunk and ignore the effects of memory consolidation which are likely to be crucial only on relatively large time scales. Mathematical description of the chunk interaction and the memory consolidation are challenging problems on their own and require individual investigation.

\section{Single Chunk Model}

\begin{figure}[t]
\begin{center}
\includegraphics[width = \columnwidth]{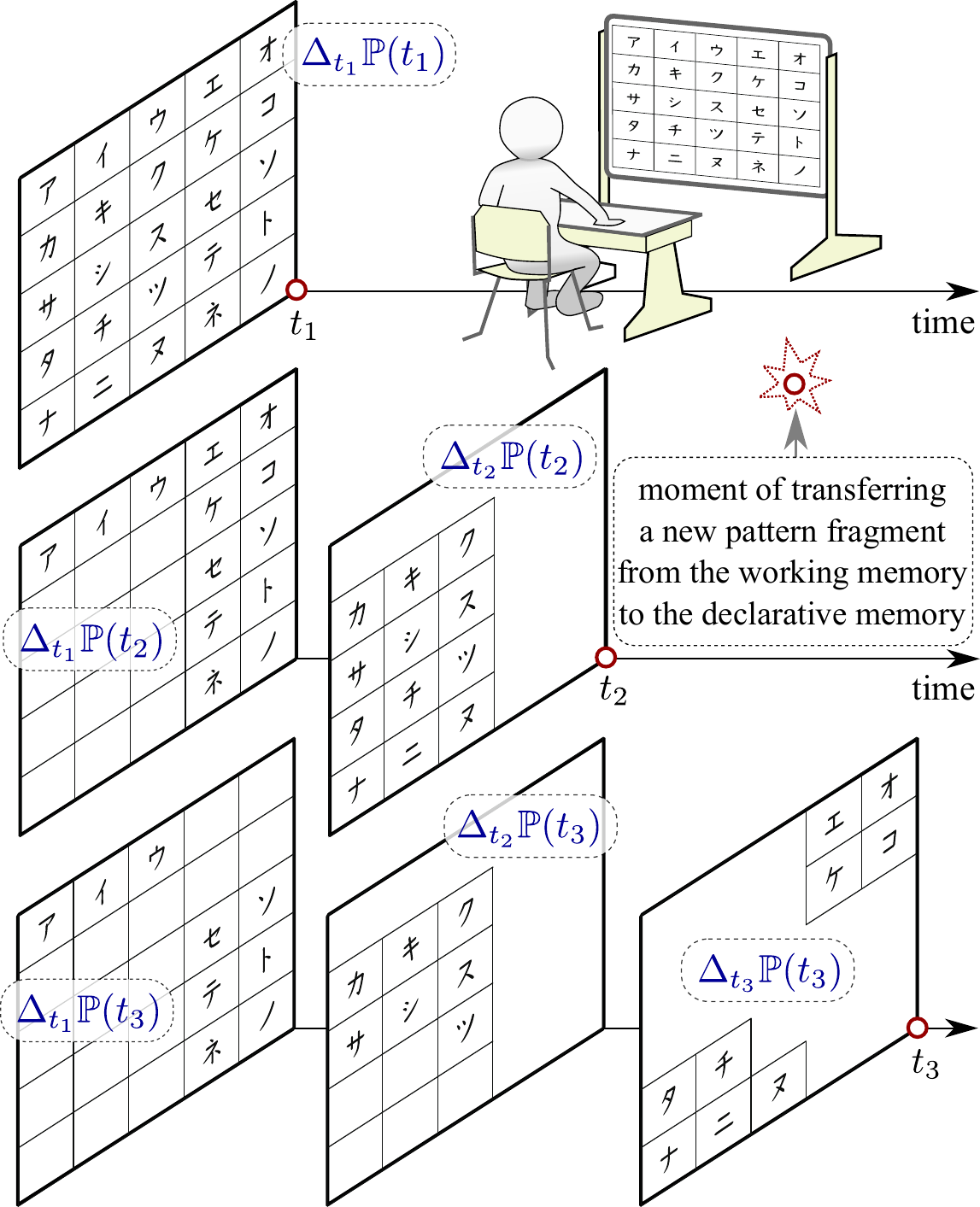}
\end{center}
\caption{Illustration of the used model for the chunk evolution as new slides are created in order to compensate the degradation of the previous ones.}
\label{Fig.1}
\end{figure}

A chunk $\mathcal{M}$ is considered to be a collection of traces $\{m(t')\}$ created in the working memory at time moments $\{t'\}$ and stored in the declarative (long-term) memory. These traces will be also called slides for reasons apparent below. Let us assume that the chunk $\mathcal{M}$ as a whole contains a certain fragment of information, a pattern $\mathbb{P}$, so all its slides retain different fragments of this pattern. 

The time evolution of the chunk $\mathcal{M}$ as a whole unit is described in terms of its strength $F(t)$, the measure of the relative volume of the pattern pieces that are retrievable at the current moment of time $t$ \cite{anderson1983spreading,anderson2007can}. 
%
%It should be noted that in the case when only one chunk is under consideration the value $B =\ln F$ is also reffed as to the basic activation level of the chunk $\mathcal{M}$.   
%
The individual evolution of the slides $\{m(t')\}$ is characterized by similar quantities $\{f(t,t')\}$ depending on the current time $t$ and the time $t'$ when the corresponding slide was created and stored. 

\subsection{Slide Formation Mechanism}

The chunk slides are assumed to be created according to the following scenario illustrated in Fig.~\ref{Fig.1}. Memory continuously looses some fragments of the pattern $\mathbb{P}$. So when at the current moment of time $t$ the chunk $\mathcal{M}$ as a whole is retrieved from the declarative memory only some its fragment $\mathbb{P}_t$ can be retrieved, which is characterized by the value $F(t)<1$. Then addition practice or learning is necessary to reconstruct the initial pattern $\mathbb{P}$. Therefore a new slide $m(t)$ to be created during this action has to contain, at least, the fragment $\mathbb{P}\setminus\mathbb{P}_t$. In principle, the pattern fragment $\Delta_t\mathbb{P}$ to be stored in $m(t)$ can include other fragments of the initial pattern $\mathbb{P}$. So in a more general case the condition $\Delta_t\mathbb{P}\cap\mathbb{P}_t \neq \emptyset$ may hold, which is worthy of individual investigation. In the present work we confine our consideration to the limit case where
\begin{subequations}\label{mod:1}
\begin{equation}\label{mod:1a}
\Delta_t\mathbb{P} = \mathbb{P}\setminus\mathbb{P}_t\,,
\end{equation}
if the current learning action is enough to create the fragment $m(t)$ containing $\Delta_t\mathbb{P}$. However, there could be a situation when the time interval $\tau$ of the learning process before the slide $m(t)$ to be transfered to the declarative memory is not enough to do this. Under such conditions we assume that before transferring the slide $m(t)$ to the declarative memory it is cut off, i.e., its capacity for new information is reduced and the saved pattern fragment meets the condition
\begin{equation}\label{mod:1b}
\Delta_t\mathbb{P} \subset \mathbb{P}\setminus\mathbb{P}_t\,.
\end{equation}
\end{subequations}
In both the cases it is reasonable to measure the capacity $C(t)$ of the new slide based on the current strength $F(t)$ of the chunk $\mathcal{M}$. Namely, in case~\eqref{mod:1a} we set 
\begin{subequations}\label{mod:c1}
\begin{align}
\label{mod:c1a}
C(t) &= 1- F(t)
\intertext{and in case~\eqref{mod:1b} the slide capacity is calculated appealing to the notion of attention $W$ paid to learning the pattern piece $\mathbb{P}\setminus\mathbb{P}_t$ during the time interval $\tau$. The following ansatz}
\label{mod:c1b}
C(t) &= [1- F(t)]\left\{1-\exp\left[-\frac{W\tau}{T(F)}\right]\right\} 
\end{align}
\end{subequations} 
is used, where $T(F)$ is the time scale characterizing the process of learning the pattern $\Delta_t\mathbb{P}$ and given by the expression 
\begin{equation}\label{mod:T}
T(F) = (\epsilon +F)^{-\alpha}(1-F)^{1-\beta}\tau_m\,.
\end{equation}
Here the scale $\tau_m$ characterizes the time interval required for the working memory to create one slide and the dependence of the quantity $T(F)$ on $F$ reflects the fact that the higher the current value $F(t)$, the less the time necessary to learn the pattern $\Delta_t\mathbb{P}$ completely, the exponents $\alpha>0$ and $\beta\in(0,1)$ specify this dependence. The parameter $\epsilon$ characterizes the duration of initial creation of the pattern $\mathbb{P}$ in the working memory.  
If we retrieve the chunk $\mathcal{M}$ immediately after this action its achievable pattern is 
\begin{equation}\label{mod:2}
\mathbb{P}_{t+\tau}= \mathbb{P}_{t} \cup \Delta_t \mathbb{P} \,.
\end{equation}
The strength $f(t,t)$ of the slide $m(t)$ created and saved just now is set equal to unity,
\begin{equation}\label{mod:3}
\left.f(t,t')\right|_{t=t'} = 1\,,
\end{equation}
which is related directly to the assumption about the reduction of the slide capacity at the moment of its creation. As time goes on, the strength of all the slides decreases and without addition learning the strength $F(t)$ of the chunk $\mathcal{M}$  as whole is written as the sum of all the slides created previously
\begin{equation}\label{mod:4}
F(t) = \sum_{t'<t} C(t')f(t,t')\,.
\end{equation}
Equalities~\eqref{mod:1}, \eqref{mod:c1}, and \eqref{mod:4} may be treated as the Bayesian approximation of the memorizing process.

\subsection{Individual Slide Dynamics}   
   
The given model assumes the slides created previously not to be affected by learning at the current moment of time. In other words, after creation their evolution is governed only by some internal mechanisms. Keeping in mind the results to be obtained let us write the equation governing the evolution of a slides $m(t')$ in the form
\begin{equation}\label{mod:5}
\frac{\partial f}{\partial t} = -\frac{a}{\tau}\,f^{\,b}
\end{equation}
where $a>0$ and $b>1$ are some parameters. Equality~\eqref{mod:3} is actually the initial condition imposed on the function $f(t,t')$. Its solution is
\begin{equation}\label{mod:6}
    f(t,t') = \left[1+\frac{(t-t')}{\tau_0}\right]^{-d} \qquad\text{for $t\geq t'$}\,,
\end{equation}
where we have introduced the new parameters $d = 1/(b-1)$ and $\tau_0 = \tau d/a$. The substitution of \eqref{mod:6} into \eqref{mod:4} yields 
\begin{equation}\label{mod:7}
F(t) = \sum_{t'<t} C(t')\left[1+\frac{(t-t')}{\tau_0}\right]^{-d}\,.
\end{equation}
It should be noted that this governing equation of the individual trace dynamics is fair similar to the mathematical model for a single trace memory proposed by \citet{wickelgren1974single}.

\subsection{Continuous Approximation}

\begin{figure}[t]
\begin{center}
\includegraphics[width = 0.9\columnwidth]{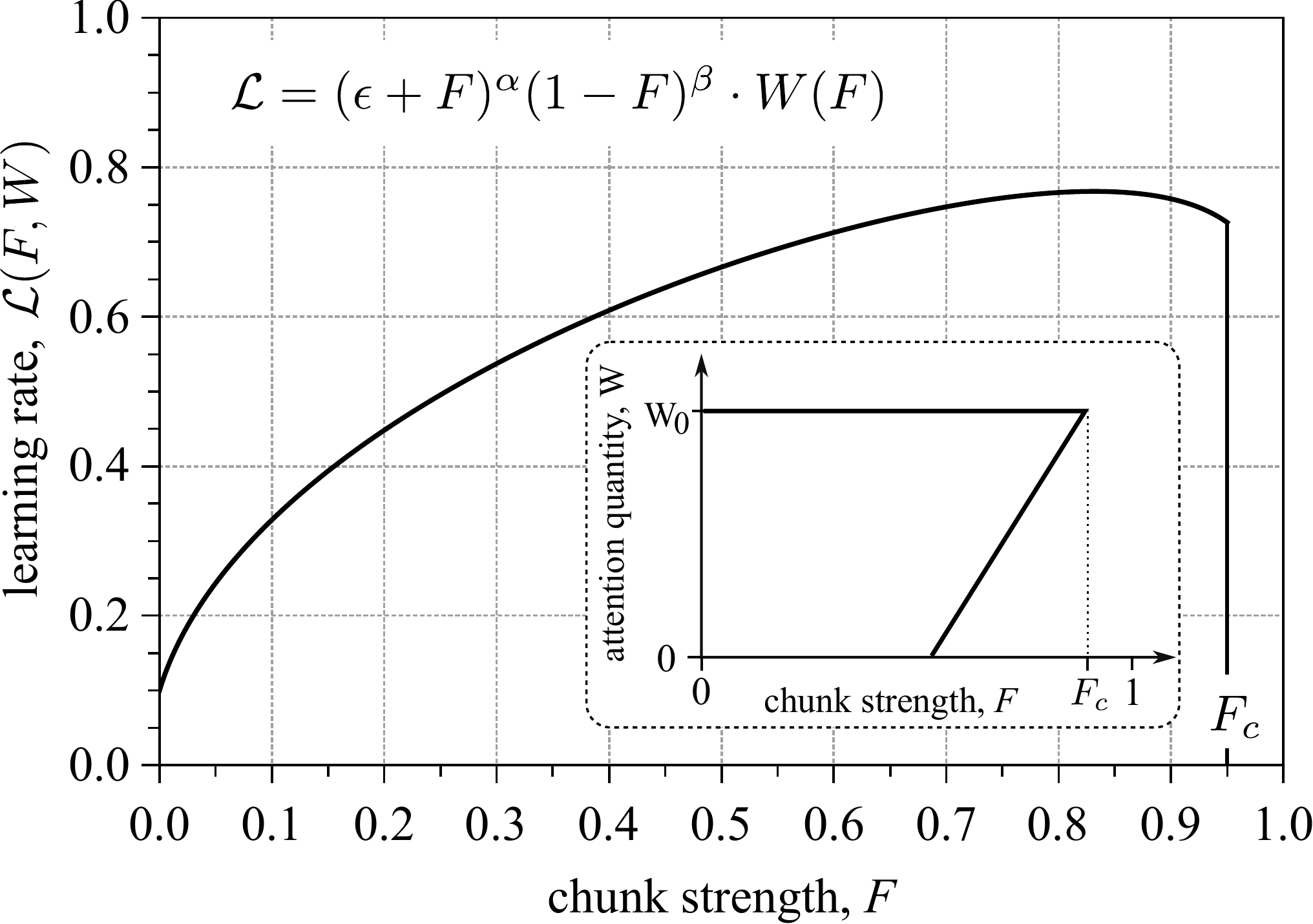}
\end{center}
\caption{The accepted ansatz for the dependence of the learning rate $\mathcal{L}(F,W)$ (in dimensionless units) on the strength $F$ of a given chunk. The used approximation for the attention quantity $W(F)$ depending on the chunk strength is shown in inset. The values of $\epsilon = 0.01$, $\alpha= 0.5$, $\beta = 0.1$, and $F_c = 0.95$ were used in constructing the present curve.}
\label{Fig.2}
\end{figure}

Expression~\eqref{mod:c1b} can be regarded as a solution $C(\zeta)|_{\zeta =\tau}$ of the equation
\begin{equation}\label{conmod:1}
\frac{dC}{d\zeta} = (1-F-C) \frac{W}{T}\\
\end{equation}
subject to the initial condition
\begin{equation}\label{conmod:2}
    C|_{\zeta=0}  = 0\,,
\end{equation}
where the values $F$ and $T$ are treated as constants, the variable $W$ describing the attention to the subject of current learning is assumed to be a smooth function on time scales about $\tau$. It enables us to represent the value of $C(t)$, Exp.~\eqref{mod:c1b}, as the cumulative result of infinitesimal increments
%Hintzman1986
\begin{equation}\label{conmod:3}
C(t) = \int dC\,,\quad\
\end{equation}
of a certain continuous process,
where 
$$
dC = (1-F)\frac{W}{T}dt\quad\text{and}\quad F(t+dt) = F(t)+dC\,.
$$
 This expression would lead exactly to formula~\eqref{mod:c1b} if the time scale $T$ were independent of $F$. However, ansatz~\eqref{mod:c1b} has been chosen rather arbitrary keeping in mind only the basic features it should possess. So we are free to replace it by the expression stemming from model~\eqref{conmod:3}. The last equality in model~\eqref{conmod:3} formally coincides with formula~\eqref{mod:4} except for the fact that the value $F(t)$ has to decrease additionally due to time evolution of temporal elements created previously. However, all the ``microscopic'' time scales, in particular, $\tau_0$ and $\tau_m$, are related directly to the interval within that a new slide is created in the working memory and, then, transfered to the declarative memory. It enables us to ignore this decrease in the value $F(t)$ during the time interval $\tau$. As a result expression~\eqref{mod:7} can be reduced to the following integral
\begin{equation}\label{conmod:4}
F(t) = \int\limits_{-\infty}^{t} [1-F(t')]\frac{W(t')}{T[F(t')]}\left[1+\frac{(t-t')}{\tau_0}\right]^{-d}dt'\,.
\end{equation}
Moreover, due to the integral 
\begin{equation*}
\int_0^t \frac1{\zeta ^d}d\zeta
\end{equation*}
converging at the lower boundary $\zeta=0$ for the exponent $d<1$ we can replace kernel~\eqref{mod:7} by the corresponding power-law kernel
\begin{equation}\label{conmod:5}
\left[1+\frac{(t-t')}{\tau_0}\right]^{-d}\quad \Rightarrow\quad \frac{\tau_0^d}{(t-t')^d} \,.
\end{equation}
After this replacement expression~\eqref{conmod:4} reads
\begin{subequations}\label{conmod:6}
\begin{equation}\label{conmod:6a}
F(t) = \int\limits_{-\infty}^{t} [1-F(t')]\frac{W(t')}{T[F(t')]}\frac{\tau_0^d}{(t-t')^d}dt'\,.
\end{equation}
or using ansatz~\eqref{mod:T}
\begin{equation}\label{conmod:6b}
F(t) = \frac1{\tau_m^{1-d}}\int\limits_{-\infty}^{t}[\epsilon+F(t')]^\alpha [1-F(t')]^\beta W(t')\frac1{(t-t')^d}dt'\,.
\end{equation}
\end{subequations}
In deriving expression~\eqref{conmod:6b} we have aggregated the ratio $(\tau_0/\tau_m)^d$ into the quantity $W$. So, first, the integral equation~\eqref{conmod:6a} contains only one microscopic time scale $\tau_m$ regarded as a certain model parameter. Second, the dimensionless quantity $W(t)$ describes the attention to the subject during the learning process. If $W=1$ then the given pattern $\mathbb{P}$ can be learned completely for a time interval about $\tau$.

Finalizing the given construction we will assume that the learning process was initiated at time $t=0$ and before it no information about the pattern $\mathbb{P}$ was available, i.e., for $t<0$ the value $F(t)=0$. In this case the integral equation~\eqref{conmod:6b} can be rewritten in the form of the following differential equation with time fractional derivative of the Caputo type 
\begin{equation}\label{conmod:final}
\tau_m^{(1-d)}\cdot{}^C{}D^{(1-d)} F = (\epsilon + F)^g(1-F)^{\gamma} W(t)\,.
\end{equation}  
It is the desired governing equation for learning and forgetting processes.   

In order to avoid some artifacts in numerical simulation we will accept an additional assumption that it is not possible to get strictly the limit value of the chunk strength $F=1$ by learning a subject. Indeed, the closer the chunk strength to unity, the more attention is necessary for a human to recognized which piece of information is unknown for him. As a result, we introduce a certain critical value $F_c\simeq 1$ such that, when the chunk strength $F$ exceeds it, $F>F_c$, a human considers the success of learning to be achieved and it attention to the learned subjects disappears, i.e.,  $W(F)=0$ for $F>F_c$. It is illustrated in Fig.~\ref{Fig.2}.

\section{Results and Discussion}

\begin{figure*}
\begin{center}
\includegraphics[width = 0.95\textwidth]{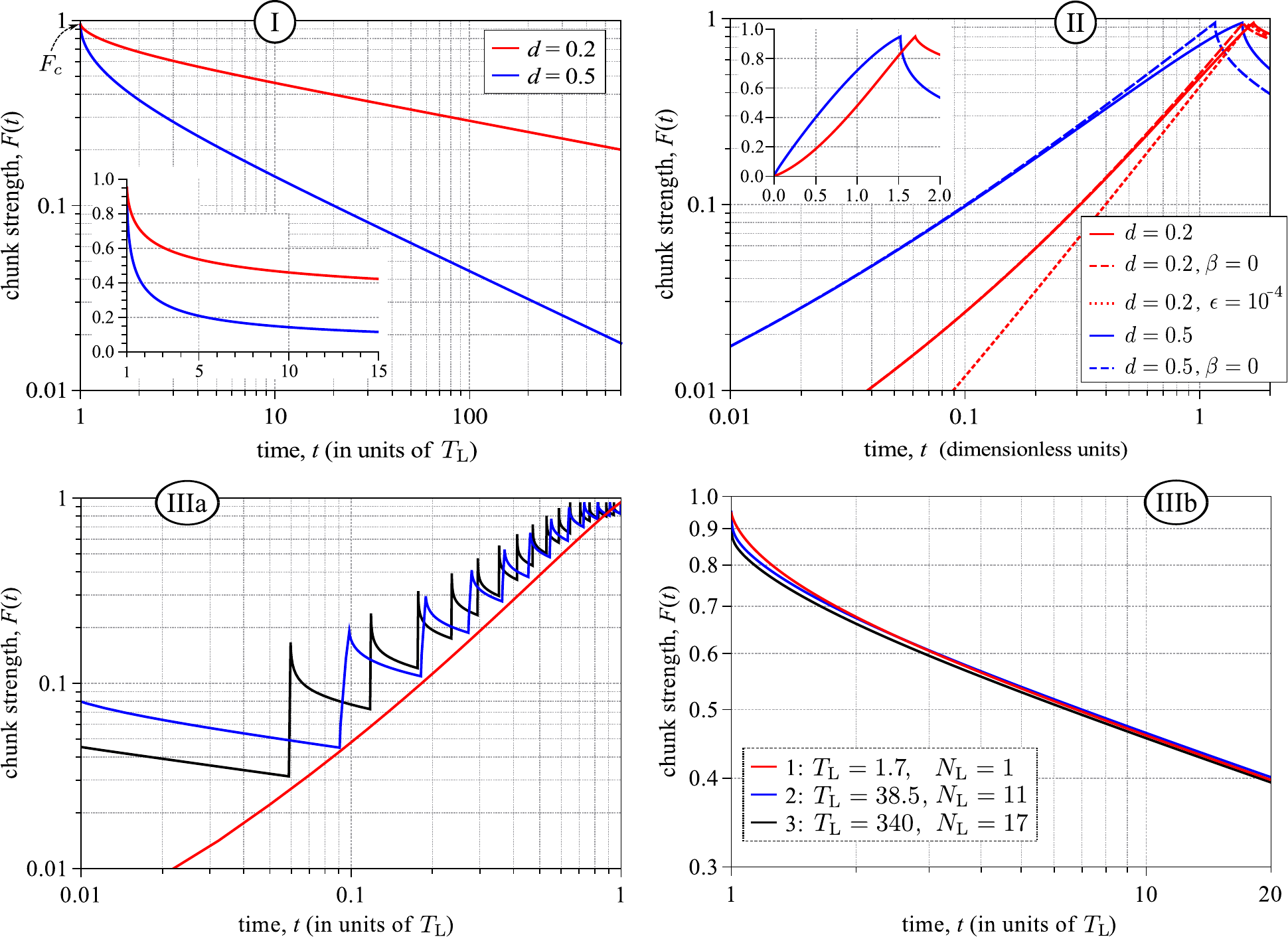}
\end{center}
\caption{Some results of numerical simulation. First, it is the dynamics of forgetting (I) and learning (II) under the ``basic'' conditions when a subject learns an unknown information continuously until he gets the local (temporal) success in time $T_L$. Second, it is the spacing effects in discontinuous learning (IIIa) and the following forgetting (IIIb). In this case the subject learns an unknown information via a sequence of trials of a fixed duration until he gets the final success at a certain moment of time $T_L$. Naturally, the longer the time spacing between two successive trials, the larger the number $N_L$ of trials and the longer the total time interval $T_L$ are necessary for this. In simulations the following parameters $\epsilon = 0.01$, $\alpha = 0.5$, $\beta = 0.1$, $F_c = 0.95$, $dt = 0.001$, and $T_\text{L}|_{d = 0.2} = 1.70$, $T_\text{L}|_{d=0.5} = 1.52$ were used as a common setup, the other individual values are shown in the corresponding plots.}
\label{Fig.SpEff1}
\end{figure*}

The characteristic features of the system dynamics were studied numerically using the explicit 2-FLMM algorithm of second order \cite{galeone2009explicit} for solving equation~\eqref{conmod:final}. Figure~\ref{Fig.SpEff1} presents some of the obtained results.

\subsection{I. Fractional Dynamics of Forgetting}

Plot~I (Fig.~~\ref{Fig.SpEff1}) shows the forgetting dynamics under the ``basic'' conditions matching the following hypothetical situation. At the initial moment of time $t=0$, a subject starts to learn continuously an unknown for him information pattern being retained in a single chunk and at time $t=T_L$ ends this process when the chunk strength gets its limit value $F_c \simeq 1$. As time goes on, the chunk strength $F(t)$ decreases, which specifies the decay of retrievable information. As should be expected, the asymptotics of $F(t)$ is of the power law and  looks like a straight line on the log-log scale plot. Naturally, in a certain neighborhood $\mathcal{Q}_L$ of the time moment $t=T_L$ this asymptotics does not hold. However, for small values of the exponent $d$ (for $d=0.2$ in Plot~I) this neighborhood is narrow and becomes actually invisible in approximating experimental data even with weak scattering.        

\subsection{II. Fractional Dynamics of Continuous Learning}

Plot~II exhibits the learning dynamics under the same ``basic'' conditions. The growth of the chunk strength $F(t)$ is visualized again in the log-log scale for various values of the parameters determining how the learning rate $\mathcal{L}(F,W)$ changes during the process (they are given in the inset). As seen, the function $F(t)$ strictly is not of the power law. However, if it is reconstructed from some set of scattered experimental points as the best approximation within a certain class of functions, a power law fit (linear ansatz in the log-log scale) may be accepted as a relevant model. It allows us to introduce an effective exponent $d_L$ of the approximation $F(t)\propto t^{d_L}$. Appealing again to Plot~II, we draw a conclusion that this effective exponent depends not only on the ``forgetting'' exponent $d$ but also on the other system parameters. Thereby, in trying to determine the set of quantities required for characterizing human long-term memory, the ``forgetting'' and ``learning'' exponents, $d$ and $d_L$, may be regarded as independent parameters.   

\subsection{III. Spacing effects}

Plots~IIIa and IIIb illustrate the found results in the case mimicking the discontinuous learning process. It again assumes a subject to start learning an initially unknown information pattern being retained in one chunk during the process, however, now he does not do this continuously. Instead, the learning proceeds via a sequence of trials of a fixed duration that are separated by some time gap (spacing) until the subject gets the success at a certain time moment $T_L$. Naturally, the longer the spacing, the longer the total time $T_L$ as well as the larger the number $N_L$ of trials required for this. So, in order to compare their characteristic properties let us renormalize the time scale in such a way that the learning process end at $t=1$ in dimensionless units, in other words, the time is measured in units of $T_L$.  In this case, as seen in Plot~IIIa and IIIb, the main characteristics of the shown processes become rather similar with respect to the dynamics of learning and forgetting. This result poses a question about optimizing a learning process by dividing it into rather short trials separated by relatively long time intervals. This effect is also called the distributed practice, an analysis of available experimental data can found, e.g., in review by \citet{cepeda2006distributed} and \citet{cepeda2009optimizing} as well. At least, within the framework of the present fair simple model an increase of the time spacing gives rise, on one hand, to the growth of the learning duration but, on the other hand, enables one to remember this information for a longer time without its considerable lost.  

\subsection{General Comments}

As far as the theoretical aspects of the present research are concerned, we note appealing to the obtained results that the multiple trace concept of memory architecture requires an individual mathematical formalism irreducible to the classical notions created in physics. In particular, even at the ``microscopic'' level dealing with slides (traces) the system dynamics is not reduced to the motion in a certain phase space but continuous generation of such phase spaces. Their interactions with one another become a key point of the corresponding theory.    

Besides, the governing equation~\eqref{conmod:final} admits the following interpretation. Its left-hand side describes ``internal'' evolution of human memory on its own, whereas the right-hand side plays the role of ``sources'' generating new elements of memory. This approach can enhance the development of human memory theory by separating the phenomena to be addressed into different categories.  

\section{Acknowledgments}
The work was supported by JSPS Grant 245404100001 (``Grants-in-Aid for Scientific Research'' Program).

%\bibliographystyle{apacite}

%\setlength{\bibleftmargin}{.125in}
%\setlength{\bibindent}{-\bibleftmargin}
%\bibliographystyle{natbib}

%\bibliography{library.bib}

\input{lubashevsky_cogsci.txt}

\end{document}